\documentclass[prb,aps,twocolumn,showpacs,floatfix]{revtex4}
\usepackage{epsfig}
\usepackage{graphicx}
\usepackage{amsmath}

\begin{document}



\title{ Surface structure in simple liquid metals. An orbital free 
first principles study.}

\author{D.J. Gonz\'alez$^1$ 
L.E. Gonz\'alez$^1$, and M.J. Stott$^2$ } 
\affiliation{$^1$Departamento de F\'\i sica Te\'orica, Facultad de Ciencias, 
Universidad de Valladolid, 47011 VA, SPAIN.}
\affiliation{$^2$ Department of Physics, Queen's University, Kingston,  
Ontario K7L 3N6 CANADA } 
\date{\today}

\date{\today}

\begin{abstract}
Molecular dynamics simulations of the liquid-vapour interfaces in  
simple {\it sp-bonded} liquid metals
have been performed using first principles
methods. Results are presented for liquid Li, Na, K, Rb, Cs, Mg, Ba, Al, 
Tl, and Si at thermodynamic conditions near their respective 
triple points, for samples of 2000 particles in a 
slab geometry. The longitudinal ionic density profiles exhibit a pronounced 
stratification extending  
several atomic diameters into the bulk, which is a feature  
already experimentally observed in liquid K, Ga, In, Sn and Hg.
The wavelength of the ionic oscillations shows a good scaling with the 
radii of the associated Wigner-Seitz spheres. 
The structural rearrangements at the interface are analyzed in terms of the
transverse pair correlation function, the coordination number and the
bond-angle distribution between nearest neighbors.
The valence electronic density profile also shows (weaker) oscillations 
whose phase, with respect to those of the ionic profile, changes from 
opposite phase in the alkalis to almost in-phase for Si. 
\end{abstract}

\pacs{61.25.Mv, 64.70.Fx, 71.15.Pd}

\maketitle

\section{Introduction}

The study of the structure of the free liquid surface has attracted much 
theoretical and experimental 
work, \cite{RowWidom-Crox,Penfold}  with 
emphasis on the possible existence of liquid surface layers.  
Although it is by now well established that ionic 
\cite{Groh-Minerva} and dielectric \cite{RowWidom-Crox,Beaglehole} liquids 
exhibit just a smooth monotonic decay from the bulk liquid density to the 
bulk vapour density, liquid surface layering appears in 
liquid crystals \cite{Ocko-MarRaton} and at the interface between a 
simple fluid and a hard wall \cite{Evans}. Understandably, special 
attention has 
been devoted to liquid metals since the early experiments suggested a 
liquid surface layered structure in Hg \cite{Barton}; computer 
simulation \cite{Rice81-83,Rice98a-99,Rice98b,Fabricius99,Walker04,GGS1} and 
theoretical studies \cite{Gomez92-94} predicted 
a significant structure, extending several atomic 
layers into the bulk liquid. These results have been recently 
corroborated by  X-ray reflectivity experiments on 
liquid Hg, Ga, In, Sn, K and Na$_{0.33}$ K$_{0.67}$. \cite{Magnussen}  

It not yet settled whether surface layering is a feature of all liquid metals 
or only characteristic of some. Rice and 
coworkers \cite{Rice81-83,Rice98a-99,Rice98b} have performed Monte Carlo (MC) 
simulations based 
on density-dependent pair potentials obtained from pseudopotentials; 
their results suggest that 
surface layering is due to the coupling between ionic and electronic 
density profiles (DP) and that the abrupt decay of the electron DP gives rise 
to an effective wall against which the ions, behaving like 
hard-spheres, stack. 
Other workers \cite{Tosatti8997} have suggested that the 
many body forces arising from 
the delocalized electrons, tend to increase the ionic surface density so that 
its coordination approaches that of the bulk. Recently, Chacon {\it et al} 
\cite{Chacon0102} proposed that surface layering may be a 
generic property of fluids at low temperature, so that the only 
requirement for an oscillatory DP is a low melting 
temperature relative to the critical temperature so as to avoid 
crystallization.

This paper reports a study on the liquid-vapour 
interface of several simple {\it sp-bonded } liquid metals
at thermodynamic conditions near their 
respective triple points. The study has been carried out by using 
the orbital-free 
{\it ab initio} molecular dynamics (OF-AIMD) method, where the forces acting 
on the nuclei are computed from electronic structure calculations, based on 
density functional theory (DFT) \cite{HK-KS}, which are performed as the MD 
trajectory is generated. 
Large samples and long MD simulation times are possible 
when interatomic pair potentials are used to describe the effective ion-ion 
interactions; however at a liquid
metal surface this approximation becomes unreliable as the electron density 
sharply drops 
from its bulk value to zero outside the surface. In these circumstances 
it is vital to allow the forces on the atoms to respond to the electron density
distribution in their vicinity. At present the best way of accomplishing 
this goal is by resorting to first-principles molecular dynamics techniques,   
where the electronic density, total energy and forces are obtained  by using the 
Kohn-Sham (KS) formulation of the density functional
theory (DFT). \cite{HK-KS}  However, the computational demands 
of these {\em ab-initio} methods, where KS orbitals are used to 
describe the electronic density and to compute exactly the electronic 
kinetic energy, grow very rapidly with system size,
and their memory requirement is also quite large. 
These considerations have posed important constrains on 
both the sizes of the systems 
studied so far, as well as the simulation times. 
Heretofore, only two KS-AIMD calculations have been performed 
on the liquid-vapour (LV) interfaces of liquid metals. 
Fabricius et al \cite{Fabricius99} have studied the LV interface in 
liquid silicon near melting by using 96 particles and a total 
simulation time of 
30 ps. Likewise, Walker {\it et al} \cite{Walker04} have studied the 
LV surface in liquid sodium near melting by a simulation 
which used 160 particles 
and lasted 50 ps. In both studies, the size limitations materialized in 
small simulation slabs ($\approx$ 12.0 \AA $\;$ thick for Si 
and 22.0 \AA $\;$ thick for Na) which may rise  
some reasonable questions about its capability to simulate a real 
macroscopic LV interface, as the small thickness makes plausible the 
existence of interactions between both sides of the slab.   
 
However, the aforementioned limitations can be partly 
overcome if the
exact calculation of the electronic kinetic energy is given up in favor of an
approximate kinetic energy functional of the electronic density. 
This is the philosophy behind the so-called orbital-free {\it ab initio} 
molecular dynamics (OF-AIMD) method \cite{Perrot-MaddenLQRT,GGLS},  
which by eliminating the orbitals of the KS formulation provides 
a simulation method where the number of variables describing the 
electronic state is greatly reduced, enabling the study of larger 
samples and for longer simulation times. 
Although the OF-AIMD method uses an explicit 
approximation for the electronic kinetic energy functional of DFT, nonetheless 
it correctly treats the forces on the ions. We stress that, although less
demanding than KS-AIMD, the OF-AIMD method is still much more costly than 
the use of pair potentials; however, recent calculations have succeeded to use  
2000 and 3000 particles to study the surface properties of several 
simple liquid metals and binary alloys, \cite{GGS1}
namely, Li, Na, Mg, Al, Si, Na$_{0.3}$K$_{0.7}$ and Li$_{0.4}$Na$_{0.6}$.

The OF-AIMD results for Na and Si, when compared with experimental data 
(as the surface tension)
and with the in principle more accurate KS-AIMD results 
(as the density profiles) for the same systems, 
can be used as a validation test of the method, as far as the study of 
metallic surface properties is concerned. As we show below in detail, several
magnitudes are very similar: the oscillating profiles, and in particular 
the wavelength of the oscillations, which is recovered exactly, the number 
of neighbors of a Si atom across the interface, and the surface tension 
of liquid Na, are all well reproduced by the OF-AIMD approach.
Further confidence in the capabilities of the OF-AIMD method in relation to
surface properties is obtained from results for metallic clusters (finite 
systems where the surface is indeed utmost important). For instance, a long 
standing, previously unexplained, anomalous variation of the melting 
temperature of Na clusters with size \cite{Haberlandetal}
has been for the first time reproduced and rationalized in terms of surface
geometry and stability \cite{Aguado} using the OF-AIMD method with the same 
kinetic energy functional used in the present work.

We do consider that reproduction of the surface properties of finite systems is
much more stringent a test than the reproduction of the surface properties of 
a semiinfinite solid surface. Nevertheless, in order to satisfy the 
requirements of one referee and to further reassess for the readers the 
capabilities of the OFAIMD method, we have performed preliminary studies of the
properties of some open metallic solid surfaces, as the $(110)$ surface of fcc 
Al and the $(10\bar{1}0)$ surface of hcp Mg.\cite{eprintAlMg} 
In both cases, experimental measurements \cite{Gobel,Ismail} 
show that surface relaxation leads to a contraction of the first 
interlayer distance, expansion of the second interlayer distance, contraction 
of the third, and so on. Moreover, the thermal expansion coefficient in the 
case of Al $(110)$ is negative, positive and positive for the first, second and 
third interdistances, while for Mg $(10\bar{1}0)$ an oscillatory behavior is 
found starting with a negative thermal expansion coefficient. 
KS-AIMD simulations for several temperatures were performed \cite{Marzari} 
for Al $(110)$ using 8 layers with 9 atoms in each one plus a vacuum 
of $8.5$ \AA, reproducing the experimental trends. 
Our OF-AIMD calculations, using the same 
simulation setup, also reproduce both the sign of the relaxations and their 
thermal behaviour.
Theoretical calculations within the quasiharmonic approximation, based on KS 
{\em ab initio} static calculations (not MD simulations),\cite{Ismail} 
were also able to reproduce the experimental behavior of Mg $(10\bar{1}0)$. 
Our OF-AIMD data again reproduce qualitatively the experimental trends, with 
both oscillatory relaxations and thermal expansion coefficients.
Further details and analysis of the data will be presented elsewhere.

In our previous OF-AIMD studies of liquid metallic surfaces \cite{GGS1} 
we first of all proved the feasibility 
of performing {\em ab initio} simulations for large systems, including one
component metals and alloys, we showed that different ordering properties 
in the case of alloys lead to substantial differences in the partial and
total density profiles, and studied the evolution of the relationship between
ionic and valence electron density profiles as the valence of the metal
is increased in the series Li, Mg, Al, Si.
Here we extend the number of systems studied, we analyze in detail the 
structure of the systems, both perpendicular and parallel to the interface, 
and complete the study of the electronic density profiles, stressing the 
evolution of these properties as the atomic valence is varied.

\section{Theory}

A simple liquid metal is treated as a disordered array of 
$N$ bare ions with valence Z, enclosed in a volume $V$, and 
interacting with $N_{\rm e}=NZ$ 
valence electrons through an electron-ion potential $v(r)$.
The total potential energy of the system can be written, within the 
Born-Oppenheimer approximation, as the sum of the direct ion-ion coulombic 
interaction energy and the ground state energy of the electronic system   
under the external
potential created by the ions, $V_{\rm ext}
(\vec{r},\{\vec{R}_l\}) = \sum_{i=1}^N v(|\vec{r}-\vec{R}_i|)$ ,

\begin{equation}
E(\{\vec{R}_l\}) = \sum_{i<j} \frac{Z^2}{|\vec{R}_i-\vec{R}_j|} +
E_g[\rho_g(\vec{r}),V_{\rm ext}(\vec{r},\{\vec{R}_l\})] \, ,
\end{equation}

\noindent where $\rho_g(\vec{r})$ is the ground state electronic density and 
$\vec{R}_l$ are the ionic positions. 

According to DFT, the ground state electronic 
density, $\rho_g(\vec{r})$,  
can be obtained by minimizing the energy functional
$E[\rho]$, which can be written 

\begin{equation}
E[\rho(\vec{r})] = 
T_s[\rho]+ E_H[\rho]+ E_{\rm xc}[\rho]+ E_{\rm ext}[\rho]
\label{etotal}
\end{equation}

\noindent
where the terms represent, respectively, 
the electronic kinetic energy, $T_s[\rho]$, 
of a non-interacting system of density $\rho(\vec{r})$, 
the classical electrostatic energy (Hartree term), 

\begin{equation}
E_H[\rho] = \frac12 \int \int d\vec{r} \, 
d\vec{s} \, \frac{\rho(\vec{r})\rho(\vec{s})}
{|\vec{r}-\vec{s}|} \, ,
\end{equation}

\noindent
the exchange-correlation
energy, $E_{\rm xc}[\rho]$, for which we have adopted the local 
density approximation and 
finally the electron-ion interaction energy, 
$E_{\rm ext}[\rho]$, where the electron-ion potential has been 
characterized by a local ionic pseudopotential  
which has been constructed within DFT \cite{GGLS}. 

\begin{equation}
E_{\rm ext}[\rho] = \int d\vec{r} \, \rho(\vec{r}) V_{\rm ext}(\vec{r}) \, ,
\end{equation}

In the KS-DFT method, \cite{HK-KS} 
$T_s[\rho]$ is 
calculated exactly by  using single particle orbitals, which requires 
huge computational effort. 
This is alleviated in the OF-AIMD approach 
\cite{HK-KS,GGLS,Perrot-MaddenLQRT} by use of an explicit but 
approximate functional of the 
density for $T_s[\rho]$. Proposed functionals consist of  
the von Weizs\"acker term, 

\begin{equation}
T_W[\rho(\vec{r})] = \frac18 \int d\vec{r} \, 
|\nabla \rho(\vec{r})|^2 /\rho(\vec{r}), 
\end{equation}

\noindent 
plus further terms chosen 
in order to reproduce correctly
some exactly known limits. Here, we have used an 
average density model \cite{GGLS}, where 
$T_s=T_W+T_{\beta}$, 

\begin{eqnarray}
T_{\beta} = \frac{3}{10} \int d\vec{r} \, \rho(\vec{r})^{5/3-2\beta}
\tilde{k}(\vec{r})^2 \\
\tilde{k}(\vec{r}) = (2k_F^0)^3 \int d\vec{s} \, k(\vec{s})
w_{\beta}(2k_F^0|\vec{r}-\vec{s}|)   \nonumber
\end{eqnarray}

\noindent
$k(\vec{r})=(3\pi^2)^{1/3} \;  \rho(\vec{r})^{\beta}$, $k_F^0$ is the Fermi 
wavevector for mean electron density $\rho_e = N_e/V$, and $w_{\beta}(x)$ is a 
weight function chosen so that both the linear response theory and 
Thomas-Fermi limits are correctly recovered. Further details  
are given in reference [\onlinecite{GGLS}].

Another key ingredient of the energy functional is the 
the local ion pseudopotential, 
$v_{ps}(r)$,  describing the ion-electron interaction. 
For each system, the $v_{ps}(r)$ has been constructed from first 
principles by fitting, within the same $T_s[\rho]$ functional, the 
displaced valence electronic density induced by an ion 
embedded in a metallic medium as obtained in a KS calculation which, moreover, 
also gives the corresponding core electronic density. 
Further details on the construction of the pseudopotential are given  
in reference  [\onlinecite{GGLS}] and we just note that 
the previous theoretical framework 
has already provided an accurate description
of several static and dynamic properties of bulk liquid
Li, Mg, Al, Si, Na-Cs and Li-Na systems \cite{GGLS,Amelie,BGGLS}.

\section{Results}

We have performed OF-AIMD simulations for the LV interfaces in the 
liquid metals Li, Na, K, Rb, Cs, Mg, Ba, Al, Tl and Si   
at thermodynamic conditions near their experimental triple points. 
For each system we have considered a slab consisting of 2000 ions 
in a supercell with two
free surfaces normal to the $z$-axis. The dimensions of the slab 
were  $L_0$ $\cdot$ $L_0$ $\cdot$  $L_z$ ($L_z$= $\alpha$ $L_0$),  
with $L_0$ and $\alpha$  chosen so that
the average number density of the slab coincides with the
experimental bulk ionic number density of the system at 
the same temperature; additional details about 
the thermodynamic states are given in 
Table \ref{SimDetails}  along with several simulation parameters. 
A further $8$ \AA $\;$ of vacuum were added both above and below the slab. 
Therefore, we are dealing with liquid slabs which are wide enough
to rule out interference effects between the two free surfaces and with 
supercells which are large enough to discard slab-slab interactions. 
Although 
the periodic boundary conditions require that a particle moving out of the cell
in the $z$-direction reappears on the other side 
of the slab, we did not observe such event during the present simulations. 
Given the ionic positions at time $t$, the electron density was expanded 
in plane waves and the energy functional was minimized with respect to 
the plane wave coefficients yielding the ground state electronic 
density, energy, and the forces on the ions, 
and therefrom the ionic positions and velocities were updated according
to Newton equations, i.e., the simulations are performed in the NVE ensemble.
For all systems equilibration runs were previously performed for a range 
between 2000-4000 configurations, depending on the system. Therefrom, the 
$N_{\rm Conf}$ ensuing configurations were those used in the evaluation 
of the slab's physical properties.

\begin{table}
\caption {Input data for the simple metals studied in this work, along with 
some simulation details. $\delta t$ is the ionic time step in ps, the
cutoff energy, $E_{\rm Cut}$, is given in Ryd, 
and $N_{\rm Conf}$ is the total number of configurations.} 
\label{SimDetails}
\begin{tabular}{cccccccccc}
\hline
& Metal &  
$\rho$ (\AA$^{-3}$) & $\;\;$ T (K) & $L_0$ (\AA)$\;\;$ & $\;\; \alpha$ & 
$\;\;\;\delta$t& $E_{\rm Cut}$ & $N_{\rm Conf}$ \\ 
\hline
& Li & 0.0445 & 470 & 28.44 & 1.95 & 0.0060 & 9.50 & 18000 & \\
& Na & 0.0242 & 373 & 33.50 & 2.20 & 0.0025 & 7.50 & 24700 & \\
& K &  0.0127 & 343 & 44.99 & 1.73 & 0.0050 & 5.25 & 25100 & \\
& Rb & 0.0103 & 315 & 48.26 & 1.73 & 0.0065 & 5.25 & 22100 & \\
& Cs & 0.0083 & 303 & 51.77 & 1.73 & 0.0050 & 4.74 & 18300 & \\
& Mg & 0.0383 & 953 & 29.90 & 1.95 & 0.0010 & 8.50 & 22000 & \\
& Ba & 0.0146 &1003 & 43.00 & 1.73 & 0.0040 & 4.95 & 21000 & \\
& Al & 0.0529 & 943 & 28.97 & 1.56 & 0.0010 & 11.25& 20000 & \\
& Tl & 0.0332 & 590 & 32.53 & 1.75 & 0.0075 & 10.50& 30000 & \\
& Si & 0.0555 &1740 & 27.41 & 1.75 & 0.0035 & 15.55& 20000 & \\
\hline
\end{tabular}
\end{table}

\begin{figure}
\begin{center}
\mbox{\psfig{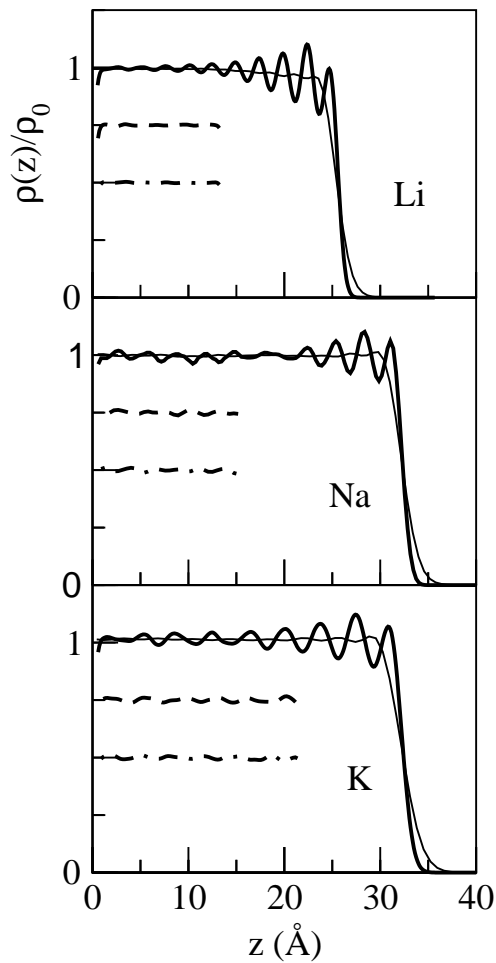}}
\end{center}
\caption{Electronic (dotted line) and ionic (full line) density 
profiles normal to the liquid-vapour interface in the Li, Na an K 
liquid metals. 
The densities are plotted 
relative to their respective bulk values. The dashed 
and dot-dashed lines are the $x$-transverse (displaced by -0.25) and 
$y$-transverse (displaced by -0.5) ionic density 
profiles.} 
\label{denproAlk1}
\end{figure}

\begin{figure}
\begin{center}
\mbox{\psfig{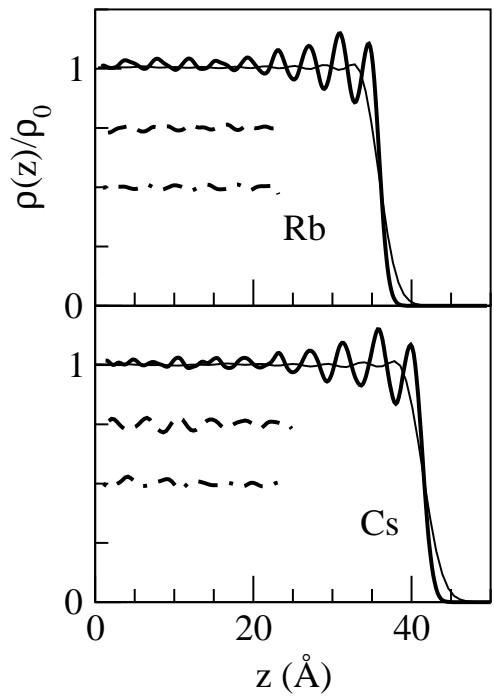}}
\end{center}
\caption{Same as the previous figure but for the 
Rb and Cs liquid-vapour interfaces.}  
\label{denproAlk2}
\end{figure}

During the simulations each slab contracted or expanded and the 
average ionic density varied  
in response to the condition of zero external pressure so that 
the ionic density in the central region of slab changed  by an amount 
ranging from $\approx$ -1.5\% in Tl to  $\approx$ 20\% in K and Si. 

\subsection{Ionic density profiles.}

\begin{figure}
\begin{center}
\mbox{\psfig{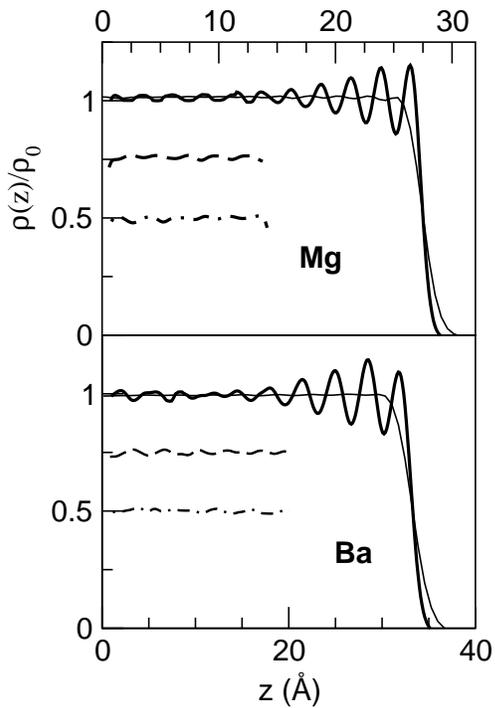}}
\end{center}
\caption{Same as the previous figure but for the 
Mg and Ba liquid-vapour interfaces. Notice the different $x$-axis scales. }  
\label{denproAlkterr}
\end{figure}

\begin{figure}
\begin{center}
\mbox{\psfig{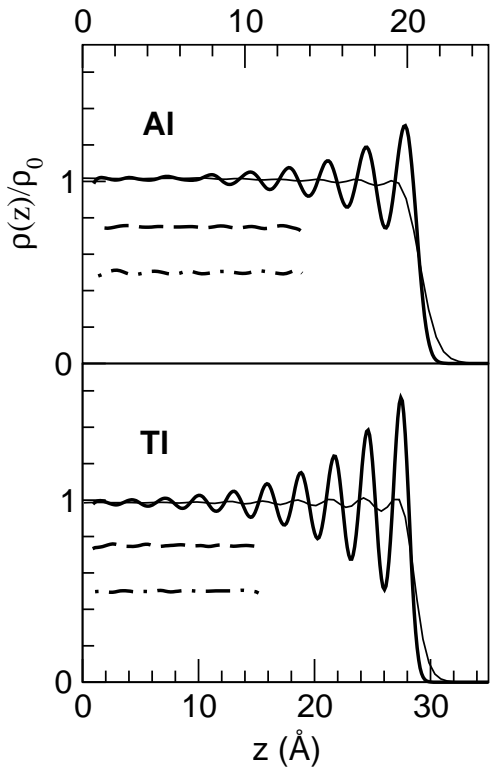}}
\end{center}
\caption{Same as the previous figure but for the 
Al and Tl liquid-vapour interfaces. Notice the different $x$-axis scales. }  
\label{denproAlTl}
\end{figure}

\begin{figure}
\begin{center}
\mbox{\psfig{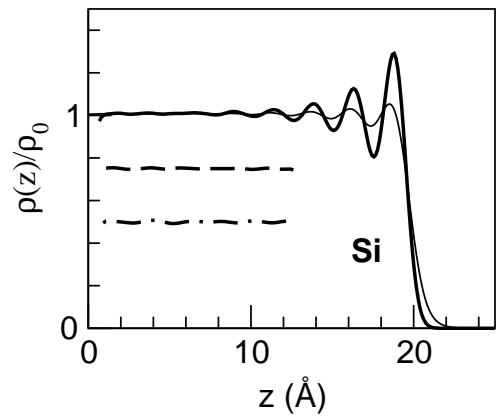}}
\end{center}
\caption{Same as the previous figure but for the 
Si liquid-vapour interfaces.}  
\label{denproSi}
\end{figure}

The longitudinal ionic DP were computed from a histogram of 
particle positions relative to the slab's center of mass, with the 
profiles from both halves of the slab being averaged, and  the  
obtained results are 
shown in figures \ref{denproAlk1}- \ref{denproSi}. 
All systems show a  stratification for at least 
four layers into the bulk liquid, a structural feature that has already been  
observed experimentally for the LV interface of 
Hg, Ga, In, Sn, K and Na$_{33}$K$_{67}$ \cite{Magnussen}.

\begin{figure}
\begin{center}
\mbox{\psfig{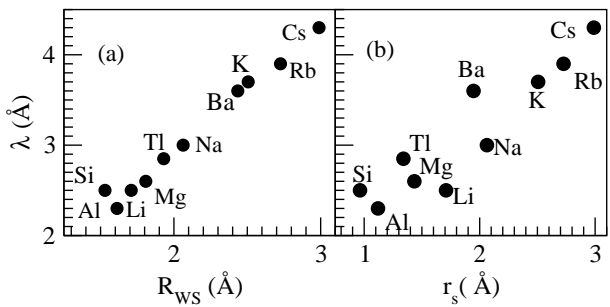}}
\end{center}
\caption{Wavelength of the longitudinal ionic oscillations 
as a function of (a) the radius of the respective Wigner-Seitz 
spheres, $R_{WS}$  
(b) the radius of a sphere which on average contains one 
electron, $r_s$. } 
\label{lambda}
\end{figure}

The wavelength, $\lambda$, of the ionic oscillations is given  
in Table \ref{Results1}, and its values scale linearly 
with the radii of the Wigner-Seitz 
spheres, $R_{\rm WS}$, while it shows no definite relationship with 
electronic parameters, like the radii per 
electron, $r_s$ (see figure \ref{lambda}).
This result suggests that the ionic oscillations are not a 
consequence of Friedel 
oscillations in the electronic density, but on the contrary they are
due to atomic stacking against the surface.
Parenthetically, we mention that the present OF-AIMD values of $\lambda$ for 
Na (3.0 \AA) and Si (2.5 \AA) exactly coincide with the values derived in 
KS-AIMD type  calculations. \cite{Fabricius99,Walker04}

Another noticeable feature of the ionic DPs refers to 
the relative amplitudes of the 
first and second oscillations near the LV interface. 
Within the group of the alkali metals, we observe that 
the amplitude of the first oscillation is smaller than that 
of the second oscillation; this feature is more marked in Li 
and is much smaller for the other alkalis. 
Conversely, for Mg we obtain a first oscillation whose amplitude is 
slightly larger than that of the second albeit this effect is reversed in Ba.
However, for Al, Tl and Si we obtain a first oscillation with an 
amplitude clearly larger than that of the second oscillation. Indeed, the 
present results suggest that as the valence of the system increases so 
does the ratio between the amplitudes of the first and second 
oscillations in the longitudinal ionic DP. 

The OF-AIMD  results for the alkalis closely agreee with  previous  
MC results of Rice {\it et al} \cite{Rice81-83,Rice98a-99,Rice98b}   
for Na, K, Rb and Cs, where the first oscillation had a  
shorter amplitude than the second. Furthermore, their MC results 
for Al, Ga, In 
and Tl \cite{Rice98b} showed a first oscillation with a larger amplitude than 
the second, which again coincides with the OF-AIMD results  
for Al and Tl, and most importantly, this trend is also visible in 
the reported experimental longitudinal 
ionic DP for Hg, Ga and In  \cite{Magnussen}. 
Likewise, we note that both the present OF-AIMD and the MC results, have 
yielded a first oscillation which is more marked in Tl than Al.  
On the other hand, the recent KS-AIMD simulation \cite{Walker04} for the LV 
interface in Na, has also produced an oscillatory longitudinal ionic DP  
with a first oscillation having an amplitude slightly larger 
than that of the second.  

Besides the correlation found between the amplitudes 
of the first and second oscillations and the valence of the system, we have 
also unveiled another correlation concerning the decaying tail 
of the ionic DPs in the LV interface region 
Specifically, when the ionic DPs are scaled in terms 
of their respective wavelength, $\lambda$, is it found that 
the decaying tail of $\rho(z/\lambda)$  is virtually the same
for all the systems with the same valence, $Z$, and moreover this tail
becomes steeper with increasing $Z$.


\begin{table}
\caption {Information about several properties of the slabs. 
$\rho$ is the ionic number density in a wide central region of the slab,  
$\Delta\rho$ is its variation with respect to the input ionic density,  
$\lambda$ is the wavelength of the longitudinal ionic oscillations,
and $\sigma$ denotes the pseudoatom size (see text). } 
\label{Results1}
\begin{tabular}{cccccccc}
\hline
& Metal &  
$\rho$ (\AA$^{-3}$) & $\Delta\rho$ (\%)  & $\lambda$ (\AA) & 
$\sigma$ (\AA) & 
$\sigma/\lambda$    \\
\hline
& Li & 0.0478 & 7.6  & 2.5 & 1.601  & 0.640 \\
& Na & 0.0272 & 12.0 & 3.0 &  1.859 & 0.620 \\
& K &  0.0152 & 19.4 & 3.7 &  2.309 & 0.624 \\
& Rb & 0.0118 & 14.4 & 3.9 &  2.458 & 0.630 \\
& Cs & 0.0089 & 7.2  & 4.3 &  2.710 & 0.630 \\
& Mg & 0.0404 & 5.5  & 2.6 &  1.500 & 0.577 \\
& Ba & 0.0162 & 11.1 & 3.6 &  2.120 & 0.589 \\
& Al & 0.0570 & 7.2 & 2.35 &  1.288 & 0.548 \\
& Tl & 0.0328 & -1.5 & 2.90 &  1.274 & 0.440 \\
& Si & 0.0666 & 20.0 & 2.5 &  1.132 & 0.453 \\
\hline
\end{tabular}
\end{table}

\begin{figure}
\begin{center}
\mbox{\psfig{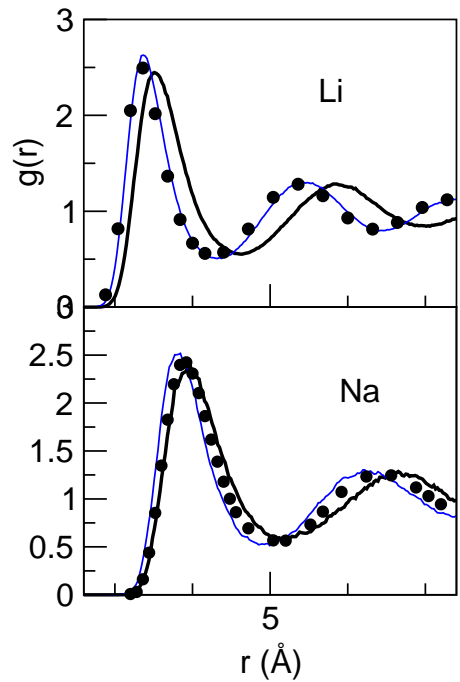}}
\end{center}
\caption{(Color online) 
Transverse pair correlation functions for selected layers 
of the Li and Na slabs, namely from the bulk (thin blue line) and   
from the outermost layer (thick line). 
The full circles stand for the bulk experimental data.}
\label{transvAlk1}
\end{figure}

\begin{figure}
\begin{center}
\mbox{\psfig{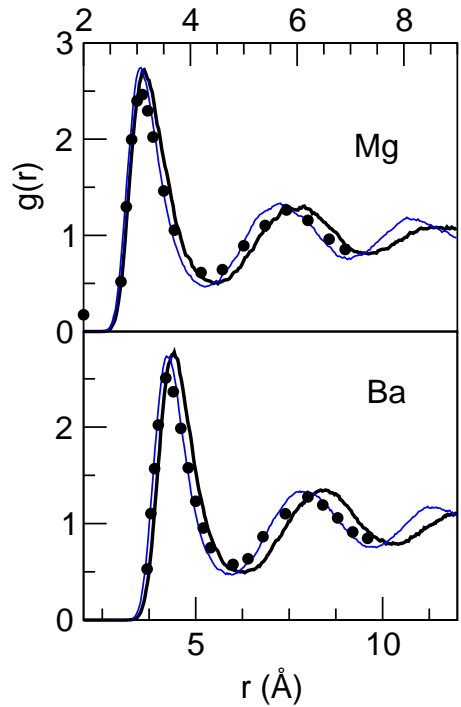}}
\end{center}
\caption{(Color online) Same as the previous figure but for liquid Mg and Ba.}
\label{transvAlkterr}
\end{figure}

\begin{figure}
\begin{center}
\mbox{\psfig{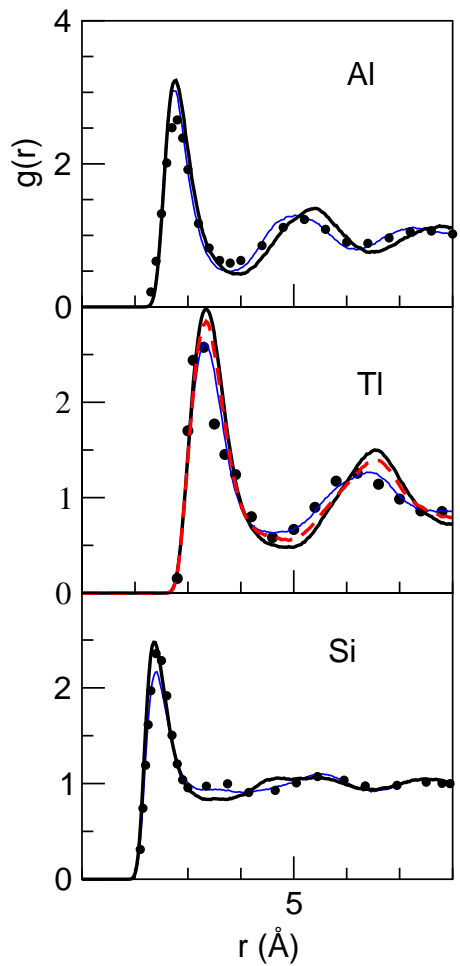}}
\end{center}
\caption{(Color online) Same as the previous figure but for liquid Al, Tl and Si. In Tl the
dashed red line is the $g_T(r)$ of the second layer.}
\label{transvAlTlSi}
\end{figure}

We have also checked that the stratification of the ionic DPs is not 
an artifact induced by the finite size of the 
simulation box. Therefore, we have computed  the transverse ionic DPs  
which, as shown in figures \ref{denproAlk1}-\ref{denproSi}, 
are more or less uniform, albeit with 
some noise which is always substantially smaller than the amplitudes 
of the oscillations in the corresponding longitudinal DP. 
As a further check on the reliability of the present calculations, we have also 
calculated the corresponding {\it bulk} pair distribution functions, $g(r)$, 
which have been evaluated within  a 
30.0 \AA $\;$ wide central section of the corresponding slab. The 
obtained results are depicted in 
figures \ref{transvAlk1}- \ref{transvAlTlSi} along with their 
experimental counterparts \cite{Wasedabook}. The small mismatch 
observed in Mg, Ba and Al may be ascribed to the aformentioned 
increase in the average ionic 
density in the central part of the slab.

\begin{table}
\caption {Details of the layers used for computing the 
transverse pair correlation functions.
$\Delta z_{OL}$ and $\Delta z_{SL}$ 
refer to the positions, with respect to the center of mass,  
of the outermost and second layers, respectively. 
$\Delta \rho_{OL}$ and 
$\Delta \rho_{SL}$ are the percent variations, with respect to the slab's bulk 
value, of the ionic number densities in 
the outermost and second layers, respectively.} 
\label{Results2}
\begin{tabular}{ccccccc}
\hline
& Metal &  
$\Delta z_{OL}$ (\AA) & 
$\Delta z_{SL}$ (\AA) & 
$\Delta \rho_{OL}$  & 
$\Delta \rho_{SL}$  & \\  
\hline
& Li & 23.6-25.7 & 21.1-23.6  & -11.3& -2.3 & \\
& Na & 29.9-32.4 & 26.9-29.9  & -6.9 &  0.0 & \\ 
& K &  29.3-32.4 & 25.6-29.3  & -8.1 & -0.5 & \\ 
& Rb & 32.8-36.1 & 28.9-32.8  & -7.2 &  0.0 & \\ 
& Cs & 37.9-41.7 & 33.6-37.9  & -9.4 &  0.0 &  \\
& Mg & 25.2-27.6 & 22.6-25.2  & -6.5 & -1.0 &  \\
& Ba & 30.2-33.4 & 26.6-30.2  & -7.7 & -0.6 &  \\
& Al & 18.7-20.8 & 16.35-18.7 & +4.0 & -1.0 &  \\
& Tl & 26.0-28.5 & 23.1-26.0  & +9.0 & -1.2 &  \\
& Si & 17.5-19.8 & 15.0-17.5  & +0.6 & -1.5 &  \\
\hline
\end{tabular}
\end{table}

\subsection{Atomic structure of the layers.}

The planar LV interfaces considered in this work are 
inhomogeneous systems along the $z$ direction, but isotropic and homogeneous  
on the $xy$ planes; consequently the
two-body distribution functions, which provide information about 
the atomic structure, 
depend on the $z$ coordinates of the two particles and on the transverse 
distance between them, i.e. $g(\vec{r}_1,\vec{r}_2)=g(z_1,z_2,R_{12})$ where
$R_{12}=\sqrt{(x_2-x_1)^2+(y_2-y_1)^2}$. The three body distribution
functions, as for instance the bond-angle distribution, also display similar
symmetry properties. 

However, such a detailed description of the structure is difficult to obtain
from the simulations, 
so below we consider several 
averaged magnitudes in which the average is taken over a given layer or slice 
of the simulation slab.
The first (or outermost) layer comprises the region from the outermost minimum
of the ionic DP to its inflection point in the decaying tail. The
other layers are located between consecutive minima of the ionic oscillations.
For a given metal, all the layers have the same width (excepting the
outermost one which is always slightly narrower) and its specific values are 
given in Table \ref{Results2}. Furthermore, in order to achieve 
meaningful comparisons we have also considered a layer with the same width and 
located in the central bulk region, which will be termed the ``center layer".

A first magnitude to analize is the average ionic density within each layer.  
Another one is the transverse pair correlation 
function $g_T(r)$, which accounts 
for the probability of finding two particles separated by a distance $r$
when both particles are inside the layer.
Then, we will consider the number of nearest neighbors (NN) and the 
distribution of angles between triplets formed by one particle and 
its two NN. Notice that in order to describe the in-plane structure 
of a layer the two previous 
magnitudes should be evaluated for those neighbours standing inside the layer.
Moroever, the particles in a layer have additional neighbors in adjacent layers
and therefore a more thorough description of the structure will require to 
account for them in evaluating the total number of NN and 
the distribution of angles.  

\subsubsection{Average ionic density.}

The calculation of the mean ionic
number density within each slice has revealed that deviations from the
bulk value are significant only in the last two slices. At the outermost slice
the relative changes in the ionic number density range from
$\approx$ -11 \% (Li) to 9 \% (Tl) whereas in the previous slice the
corresponding values are substantially smaller (see Table \ref{Results2}).
Note that the average ionic density in the outermost layer decreases in all the
alkalis and alkaline-earths, and increases for Tl, Al and Si.

\subsubsection{Transverse pair correlation functions.}

In all the systems studied, excepting Tl, the change in $g_T(r)$ from 
the bulk to the surface occurs rather abruptly at the outermost layer, 
because at the second layer its $g_T(r)$ already coincides with that of 
the center layer, which in turn is practically identical to the bulk $g(r)$. 
As for Tl, its LV interface is the most structured one, with  
large oscillations which may indicate 
an important 
influence of any 
given layer on the properties of its surrounding layers. In fact, the change 
in $g_T(r)$ from the bulk to the surface is gradual in the case of Tl.

The changes undergone by the average ionic density (Table \ref{Results2}) 
are mirrored by deviations of the associated $g_T(r)$ with respect to the 
bulk $g(r)$, as evinced by 
figures \ref{transvAlk1}-\ref{transvAlTlSi}, which show the
transverse pair correlation function $g_T(r)$ at the outer layer.
For both alkalis and alkali-earths, the outermost $g_T(r)$ is displaced
towards greater $r$-values, in clear correlation with a decreased average
ionic density. 
Conversely, the outermost $g_T(r)$ for Al, Tl and Si practically preserve
the same main peak's position as in the
bulk while slightly increasing its height. Notice that in these latter three
systems, the ionic number density at the outermost layer did increase with
respect to the slab's bulk value.

Conceivably, it might appear that for the alkalis and alkaline-earths 
those changes
in $g_T(r)$ just amount to an expansion of the system, leading to an increased
NN's distance. In the case of Al and Tl the maximum of $g_T(r)$
does not change position, but the first minimum moves somewhat to the right. 
The most interesting situation occurs for Si, where apparently 
some atoms move from the position of a small bump
just after the main peak of $g(r)$ to a distance slightly smaller than the 
position of the second maximum, yielding a double-second-peak structure in 
the $g_T(r)$ of the outermost layer. All these changes induced by the 
layering of the interface will be thoroughly analyzed in the following 
section.  

\subsubsection{In-plane neighbors.}

The number of NN, or coordination number (CN), is usually defined
as the average number of neighbors within a distance $r_{\rm max}^{2d}$, 
identified
as the position of either the first minimum of the pair correlation function 
$g_T(r)$, or that of the radial distribution function $G_T(r)$ (which for 
these quasi-bidimensional layers is proportional to $rg_T(r)$ and the average
ionic density) of the layer considered. 
This latter criterion has been used to calculate the results of 
Table \ref{2d-coonum} although similar 
trends are found when using the former one. In the alkalis and 
alkaline-earths there is a competition between the increase of 
$r_{\rm max}^{2d}$
at the outermost layer and the corresponding decrease of the average 
ionic density.
In all cases the latter factor is stronger and we find that the CN,  
$n_{\rm 2d}$
changes from about $5.2$ in the center layer to about $4.7$ in the outermost 
one. In Al and Tl both the average ionic density and $r_{\rm max}^{2d}$ 
increase leading 
to a somewhat increased CN from the center to the surface. In Si 
both factors change very little and we find $4$ NN (in-plane) 
both for the center layer and for the outermost one.

\begin{table}
\caption{Number of neighbors within a layer $n_{\rm 2d}$ for the systems 
studied.}
\begin{tabular}{ccccc}
\hline
Metal & \multicolumn{4}{c}{$n_{\rm 2d}$} \\
 & Layer 1 & Layer 2 & Layer 3 & center layer \\
\hline
Li & 4.8 & 5.3 & 5.3 & 5.3 \\
Na & 4.5 & 5.2 & 5.2 & 5.2 \\
K  & 4.6 & 5.3 & 5.1 & 5.1 \\
Rb & 4.6 & 5.1 & 5.1 & 5.1 \\
Cs & 4.7 & 5.2 & 5.2 & 5.2 \\
Mg & 4.9 & 5.1 & 5.1 & 5.1 \\
Ba & 4.8 & 5.3 & 5.3 & 5.3 \\
Al & 5.1 & 4.9 & 4.8 & 4.8 \\
Tl & 5.5 & 5.5 & 5.2 & 5.1 \\
Si & 4.1 & 3.9 & 4.1 & 4.0 \\
\hline 
\end{tabular}
\label{2d-coonum}
\end{table}

Notwithstanding the validity of these results it must be recognized that there
is a certain degree of arbitrariness in the definition of when a particle
is a ``nearest" neighbor of another. Some years ago McGreevy and 
coworkers \cite{McGreevy86} outlined 
a proposal for a less arbitrary determination of the CN 
in computer simulation studies.
Their idea is to rewrite the radial distribution function as the sum of the
partial functions,  $G_T(r)=\sum_{i=1}^{\infty} G_T^i(r)$, where $G_T^i(r)$ is
the pair distribution function corresponding to the $i$-th neighbor, which 
indicates the distribution of distances from one particle to its $i$-th 
neighbor. By analyzing the way the different neighbors are distributed, 
it is possible to discern 
whether they are inside the first peak, outside it or 
in between, and 
therefore it provides a more precise picture of the meaning of the CN.  
In the present context this idea is even more useful, since
by looking at the change of $G_T^i(r)$ across the LV interface we 
can better describe the rearrangement of the particles induced by the 
layering of the interface.

\begin{figure}
\begin{center}
\mbox{\psfig{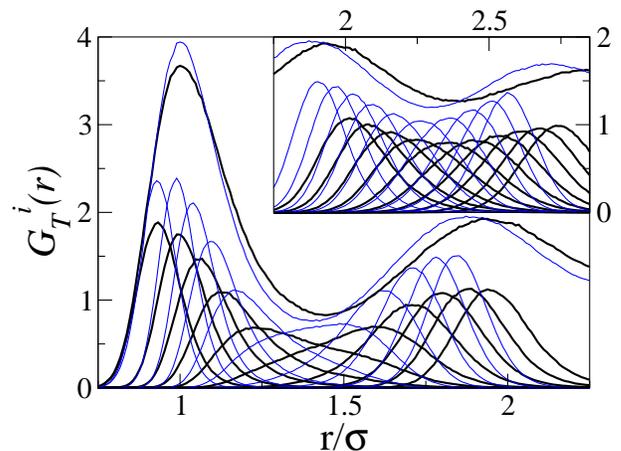}}
\end{center}
\caption{(Color online) Decomposition of the pair distribution functions for Li 
in terms of those
corresponding to the first, second, $\cdots$ ,up to the 20th neighbor.
Thick lines: outermost layer, thin blue lines: center layer. 
Also shown are the associated pair correlation functions, $g_T(r)$ 
(scaled so as to fit into the graph).}
\label{deconvLi}
\end{figure}

\begin{figure}
\begin{center}
\mbox{\psfig{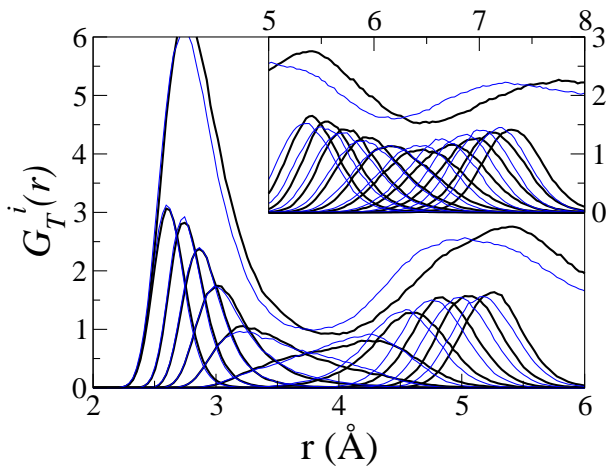}}
\end{center}
\caption{(Color online) Same as the previous figure but for Al.
}
\label{deconvAl}
\end{figure}

\begin{figure}
\begin{center}
\mbox{\psfig{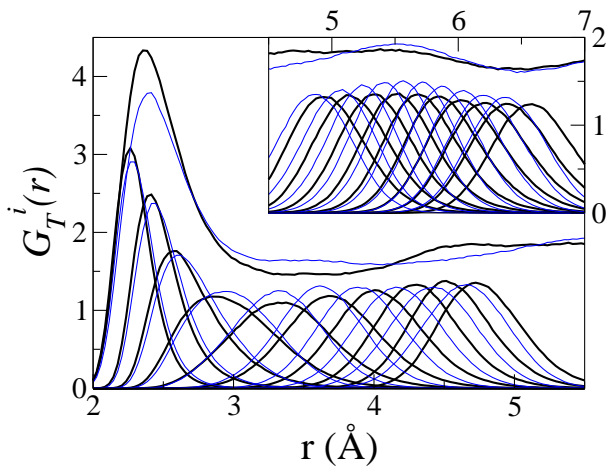}}
\end{center}
\caption{(Color online) Same as the previous figure but for Si.
}
\label{deconvSi}
\end{figure}

In figures \ref{deconvLi}-\ref{deconvSi} we show this decomposition 
of $G_T(r)$ for Li (which stands as 
a representative example of the alkalis and alkaline-earths, 
because all of them show the same behavior), Al and Si. 
Note that in all cases the amplitudes of $G_T^i(r)$ follow closely the shape 
of the pair correlation function, $g_T(r)$, of the layer considered.

Let us consider first the case of Li at the center layer. It is found that the 
neighbors up to the fourth are well below the first peak of $g_T(r)$, and the
fifth neighbor, despite being more spread out, can reasonably be assigned to 
the set of ``nearest" neighbors. The sixth one, however, is clearly distributed
between the first and second peaks of $g_T(r)$. The second peak is 
spawnned by 
part of the sixth neighbor, the neighbors number 
7 to 15 and again partly the 16th one. From this analysis one could assign
$5.4$ and $10.1$ neighbors to the first and second coordination shells 
respectively. In the case of the outermost layer we find 4 neighbors well 
inside the first peak of $g_T(r)$, whereas the fifth and sixth neighbors are 
very spread out, both being shared by the first and second peaks of $g_T(r)$.
The rest of the second peak is due to neighbors number 7 to 15 and part of the
16th one. We can therefore ascribe values of $5$ and $10.5$ to the first and
second CNs.
In order to separate out the effects of the decrease of the average ionic 
density, the comparison between
the $G_T^i(r)$ of the center layer and that of the outermost one is made
for Li by rescaling the distances by $\sigma$, which denotes the position of 
the maximum of $g_T(r)$. In this way we observe that the expansion is not
uniform: the peak positions of $G_T^1(r)$ to $G_T^3(r)$ remain unchanged 
(in units of $\sigma$) but from the 4th neighbor onwards there is more than
uniform expansion and already the 6th neighbor in the outermost layer is
located at the position of the 7th one in the bulk, while the 12th neighbor in
the outermost layer is located at the position of the 14th one in the center.

For liquid Al the rearrangement of atoms at the interface is very small. 
The first five neighbors hardly change, whereas starting at the sixth neighbor
a small expansion occurs, followed again by contraction so that the positions
of the 15th neighbors again coincide. Analyzing the distribution of atoms 
under the first and second peaks, the CNs for the first
and second shell would be $5.0$ and $9.5$ in the center and $5.2$ and $10.3$ at 
the surface.  

In Si we find that the small bump after the main peak of the bulk $g(r)$ 
is due to the particular distribution of the fifth and sixth neighbors, the 
first four being clearly located under the main peak, which leads 
to a CN of $4.0$ atoms, both in the center and at the surface. 
When one comes to the outermost layer the heights of $G_T^5(r)$ and $G_T^6(r)$ 
are depleted, whereas
the following neighbors up to the 16th attain very similar distributions
although of course at increasing distances, yielding this way the double-peak
structure of the second peak of $g_T(r)$. Therefore the previous interpretation
of some atoms moving from the bump to the second peak is not correct; 
the position of the fifth neighbor is exactly the same in the surface and in 
the bulk, and starting from the sixth neighbor a weak expansion does occur, 
but it amounts to one atom only at large distances: the 18th surface neighbor 
is located at the position of the 19th bulk neighbor.

\subsubsection{In-plane bond-angle distribution functions.}

\begin{figure}
\begin{center}
\mbox{\psfig{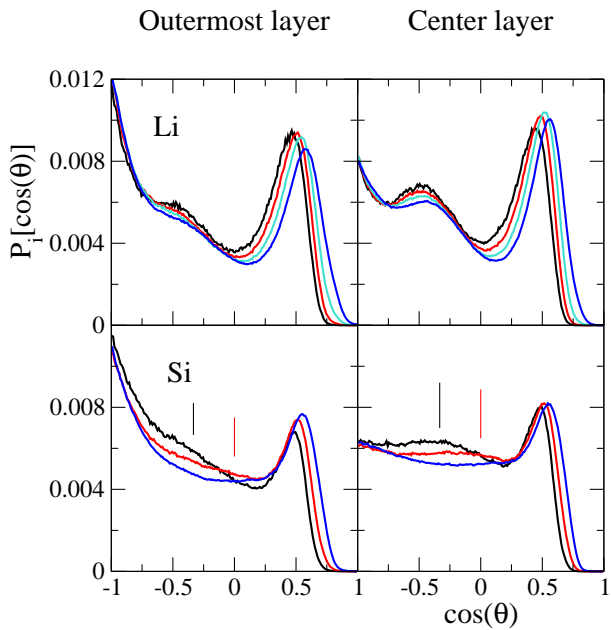}}
\end{center}
\caption{(Color online) Bond-angle distribution functions for Li and Si at different layers 
of the interface. The number of neighbors included in the calculation 
increases from 2 towards the coordination number (5 for Li, 4 for Si)
as the position of the peak near $\cos \theta = 0.5$ moves to the right. 
The vertical lines indicate
the cosines of the tetrahedral angle (109.5 degrees) and of 90 degrees.}
\label{LiSi_inplane}
\end{figure}

The usual way to calculate the bond-angle distribution function is to choose 
an atom and compute the angle between the lines joining this atom and two of
its NN, which would mean that they are separated less 
than $r_{max}^{2d}$ from the first one. However, following the ideas of 
the previous section we compute the angle between the lines joining
the central atom and any two atoms taken from its first $i$ neighbors. 
It is interesting to analyze the evolution of the bond-angle distributions, 
$P_i(\theta)$, as more NN are included by starting at 
$i=2$ and up to the integer closest to the CN. 
Figure \ref{LiSi_inplane} shows the obtained 
results for liquid Li and Si. 

Starting with Li, we observe that in the center layer the distribution of 
cosines of angles peaks near $0.5$, $-0.5$ and $-1$ (angles 60, 120 and 
180 degrees) suggesting that the local environment is mainly hexagonal, even 
though the CN is around 5 (in a penthagonal environment the
peaks would occur at $0.309$ and $-0.809$ very far from the observed ones).
As we increase the number of neighbors included in the calculation the peak
around 60 degrees moves towards smaller angles, whereas the one near 120 
degrees becomes fuzzier while hardly changing its position. The picture
in the outermost layer is rather similar though the peak near 120 degrees is
reduced to a shoulder in favor of an enhanced peak at 180 degrees.

The case of Si, with a more open structure, shows a much stronger variation in
the bond-angle distributions as more neighbors are included in the
calculation. Starting with the center layer, we observe a clear peak near 60 
degrees, which moves towards smaller angles if more neighbors are included.
However, if only the first two neighbors are included a faint feature also
appears near the tetrahedral angle, when including up to the third neighbor, 
this features moves to an angle near 90 degrees, and if we consider all 
the neighbors up to the fourth, the feature completely disappears.
The variations brought about at the surface layer are similar as those 
observed in 
Li: we find an enhanced peak at 180 degrees, while the small features observed 
at the center layer for intermediate 
angles show up as shoulders, which also move their position from thetrahedral 
to 90 degrees and to disappearance when including 2, 3 or 4 neighbors in the 
computation.

\subsubsection{Total number of neighbors.}

To analyze the local 3-dimensional structure of the atoms at the
different layers, we have computed the corresponding $z$-dependent 
CN, $n(z)$, by counting the average number of neighbors (whatever be their
position) within a distance $r_{\rm max}^{3d}$ taken as the 
position of the 
first minimum of the bulk pair distribution function (proportional 
to $r^2g(r)$).
Our results, which are summarized in
Table \ref{Results3}, show that for most of the slab $n(z)$ remains
invariable and close to that of the bulk system and
only very close to the interface, namely starting around the second
outer maximum, $n(z)$ begins to decrease. Aside from liquid Si, all
the other systems have an average bulk CN, 
$n(z_B)$ $\approx$ 12, which is reduced by $\approx$ 1/3 at their respective
outer maximum, namely $n(z_{OM})$ $\approx$ 8.  Liquid Si has a noticeably
smaller $n(z_B)\approx 6$, decreasing to $\approx 4.5$ at the surface,
because of some remmants of covalent bonding 
in the liquid, which induce an open structure with an experimental value of
around 5.7 \cite{Takeda}- 6.4 \cite{Waseda} neighbours.
For comparison, we note that the KS-AIMD calculations of
Fabricius {\it et al} \cite{Fabricius99} for liquid Si yielded 
$n(z_B)$ $\approx$ 6.4 which reduced to 5.3 at the outermost maximum.
Indeed, this decrease in CN is a well known feature at
surfaces, produced by the lack of atoms outside the interface. In the
next section we will discuss if (and how) the remaining neighbors 
redistribute in this outer part of the system.

As for the structural rearrangements induced by
the interface, Fabricius {\it et al} \cite{Fabricius99} 
have proposed to quantify those changes 
by comparison with an ideally terminated surface obtained by cutting abruptly
the slab in the central region, i.e. at $z=0$. Then, we evaluate $n(z)$ at
a distance $z_d$  which is approximately the distance
between the outermost maximum and the inflection point in the respective 
ionic DP. The obtained values are shown in
Table \ref{Results3} and they are slightly smaller than
those at the outer maximum which suggests that the surface structural
rearrangement induces some minor increase of the CN of
an ideally terminated surface.

\begin{table}
\caption {Values of the $z$-dependent coordination number, $n(z)$ at diferent
$z$-values along the slab.
$n(z_B)$ is the average value in a wide region around the center of the slab.
$z_{OM}$ and $z_{SM}$ are the positions (in \AA) of the outermost and second 
maximum respectively.
$z_{d}$ is the distance (in \AA) between the outermost maximum and the
inflection point at the decaying ionic density profile, and
$n(z_{d})$ is defined in the text.}
\label{Results3}
\begin{tabular}{cccccccccc}
& Metal &  $n(z_B)$ &  $ z_{OM}$ & $n(z_{OM})$  &
$ z_{SM}$ & $n(z_{SM})$  & $z_{d}$ & $n(z_{d})$  \\
\hline
& Li & 11.85 & 24.7 & 7.9 & 22.4 & 11.3 & 1.05 & 7.8 & \\
& Na & 11.60 & 31.1 & 7.9 & 28.3 & 11.2 & 1.30 & 7.6 & \\
& K  & 11.90 & 30.9 & 7.9 & 27.4 & 11.5 & 1.45 & 7.8 & \\
& Rb & 12.00 & 34.5 & 8.1 & 30.9 & 11.7 & 1.55 & 7.8 & \\
& Cs & 12.00 & 39.9 & 8.0 & 35.8 & 11.5 & 1.80 & 7.8 & \\
& Mg & 11.50 & 26.4 & 7.9 & 23.9 & 11.3 & 1.25 & 7.7 & \\
& Ba & 12.00 & 31.8 & 8.1 & 28.4 & 11.6 & 1.60 & 7.9 & \\
& Al & 11.30 & 18.7 & 7.9 & 16.35& 11.2 & 1.05 & 7.6 & \\
& Tl & 11.90 & 27.5 & 8.2 & 24.6 & 11.9 & 1.0  & 7.6 & \\
& Si& 5.9  & 18.8 & 4.5 & 16.3 & 5.8  & 1.15 & 4.3 & \\
\end{tabular}
\end{table}

\subsubsection{Total bond-angle distribution functions.}

\begin{figure}
\begin{center}
\mbox{\psfig{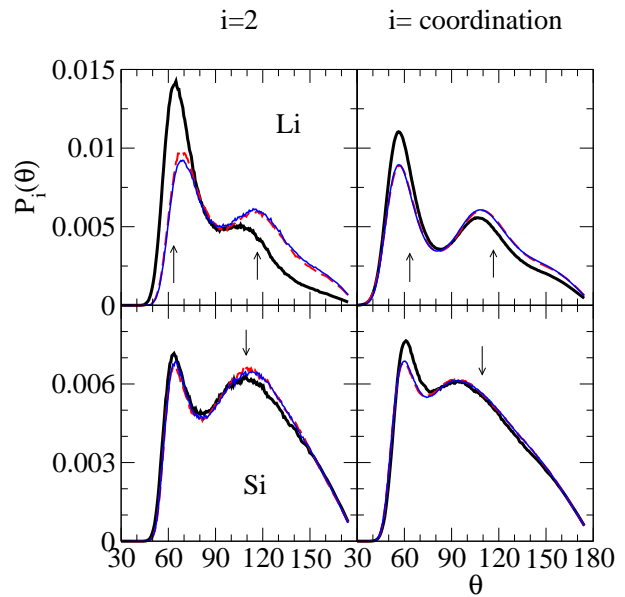}}
\end{center}
\caption{(Color online) Total bond-angle distribution functions for Li and Si.
Thick line: outermost layer. Dashed red line: second layer. Thin blue line:
center layer. 
The number of neighbors included in the calculation is 2 for the left
panels and the coordination number 
for the right panels.
The arrows indicate the icosahedral angles (63.4 and 116.6 degrees) for Li 
and the tetrahedral angle (109.4 degrees) for Si.
}
\label{LiSi_inwide}
\end{figure}

Again, instead of the usual way of defining a NN as one separated
by at most $r_{\rm max}^{3d}$ from the central one, we compute the bond angles
between a particle and two of its first $i$ neighbors, with emphasis on $i$ 
being either 2 or the integer value closest to the (3-dimensional) CN.  
Figure \ref{LiSi_inwide} depicts the results obtained for 
the bond-angle distribution functions in Li and Si. Specifically, we show the 
results obtained by counting only the first two neighbors in the 
calculation as well as 
those obtained by considering the corresponding CN, which 
amounts to 8 neighbors for Li at the outermost layer and 12 at the inner ones,
and 4 neighbors for Si at the surface, and 6 at the second and center layers.

In both systems we find that when 
the calculation includes all the NN 
(the full CN), then the bond-angle distribution 
at the surface closely resembles that at the inner layers, including the bulk; 
however, we stress that this result relies on taking the CN associated 
to each layer, otherwise the distributions would peak at somewhat 
different angles. The more probable angles are somewhat smaller than those
reflecting a perfect icosahedral environment in Li, and around 60 degrees and 
a somewhat smaller angle than the tetrahedral one in Si.

On the other hand, if the calculation includes only the first two 
neighbors, then the obtained  bond-angle distribution describes a 
situation where the disorder induced by the distance is somehow at its 
minimum level, thus reflecting the ``most local" bond distribution.
The results shown on the left panels of figure \ref{LiSi_inwide} provide 
evidence that in the case of Si, the surface has practically no influence on 
the bond-angle distribution which now shows a peak around 60 degrees 
and another one at exactly the tetrahedral angle.
For liquid Li we find peaks close to the icosahedral angles in 
the inner layers, but at the surface we observe a 
redistribution of bonds with an increased number of bonds at around 
60 degrees along with an important depletion 
at larger angles. However, the peak's positions do not change appreciably 
and remain around the icosahedral angles.

\begin{figure}
\begin{center}
\mbox{\psfig{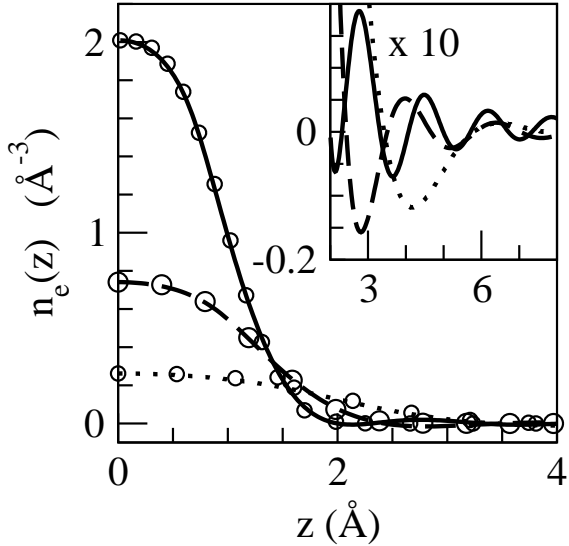}}
\end{center}
\caption{ Valence pseudoatomic densities 
for K (dotted line), Mg (dashed line) and Si (full line); the inset  
depicts its shape (multiplied by a factor of ten) for larger distances.
The open circles represent the analytical model proposed 
in the text.}
\label{nz-val-KMgSi}
\end{figure}

\subsection{Electronic structure.}

Figures \ref{denproAlk1}-\ref{denproSi} show the calculated  
self-consistent valence electronic DP which also exhibit some 
oscillations although with much smaller amplitudes than those of 
the associated longitudinal ionic DP.  
However, the most remarkable feature refers to  
the relative phases between the longitudinal ionic and valence electronic DP.  
According to Figures \ref{denproAlk1}-\ref{denproSi} all situations are 
allowed with the relative phase evolving from an opposite one, as it 
happens for all the alkalis, to being almost in-phase for Si. 

This is an unexpected feature because the scant previous studies on the 
longitudinal ionic and the valence electronic DP \cite{Rice98a-99,Jesson} 
had always yielded an opposite phase. Specifically, the    
Monte Carlo calculations of Rice and coworkers \cite{Rice98a-99} for 
the LV interface in liquid Na, Cs and Ga produced longitudinal 
ionic and valence 
electronic DP standing in nearly perfect opposite phase. 
This feature was rationalized in terms of a  
competition between the valence electronic kinetic
energy contribution, which gets smaller values by damping 
the oscillations, and the interaction term between electrons 
and ions, (mainly its coulombic part), 
which being attractive takes smaller values for in-phase oscillations. 
Furthermore, the OF-AIMD simulations by Jesson and Madden \cite{Jesson} 
for the structure of the liquid-solid interface in Al also gave a 
longitudinal valence electronic DP standing in a opposite phase with the ionic 
one in both the crystalline and liquid phases. This behaviour was now  
explained in terms of the interaction, embodied in the pseudopotential, 
between valence and core electrons, which tends to exclude the valence 
electrons from the ionic positions.

In fact, the assumption of an opposite phase between the 
longitudinal ionic and valence electronic DP has been widely assumed; however 
a closer scrutiny, based on the present calculations, reveals some 
weak points on the previous explanations. First, our results for the 
valence electronic kinetic energy show that it takes very small 
values (less that 5\% for all systems in this work) in comparison   
with those of the valence electron-ion interaction term; therefore  
the idea of assigning to the valence electronic 
kinetic energy any relevant role in  
establishing the phase of the valence electronic DP'oscillations appears 
baseless.  Second, all the systems considered in this work, 
have $s$-type valence electrons which in some 
cases (Al, Tl and Si) are also joined by $p$-type electrons. Taking into 
account that the $s$-type valence electronic density attains  
a maximum value at the ionic positions, it now appears rather surprising to 
find systems, specially those with $s$-type valence electrons only, in 
which both profiles were not in-phase. 
Consequently, we have undertaken several tests in order to ascertain 
the mechanisms underlying the 
phase-shift between the longitudinal ionic and valence electronic DP. 

\begin{figure}
\begin{center}
\mbox{\psfig{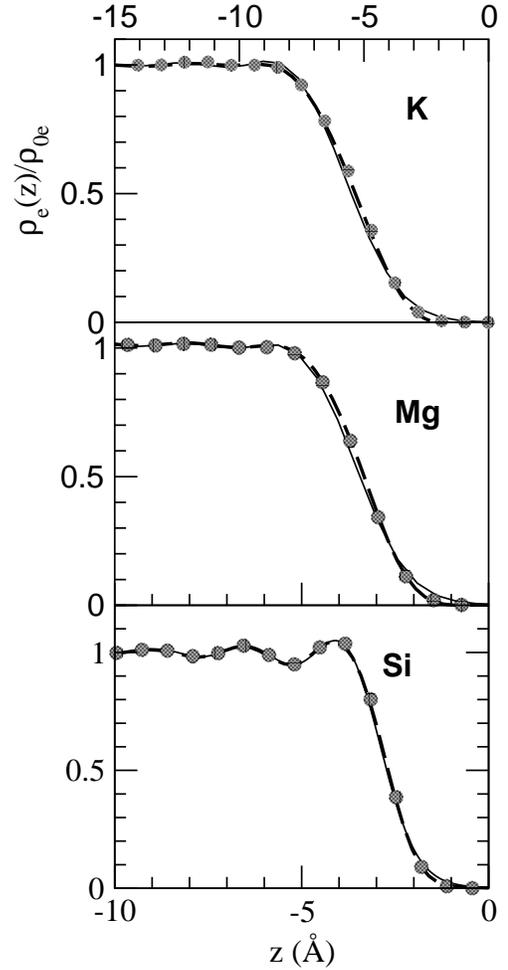}}
\end{center}
\caption{Valence electronic density profiles (relative to the 
bulk average value) for 
the K, Mg and Si liquid-vapour interfaces. 
The continuous line is the OF-AIMD result, the dashed 
line represents a linear superposition of displaced 
densities and the grey circles are a linear superposition of the 
model densities, without Friedel oscillations. }
\label{denelecKMgSi}
\end{figure}

First, we have constructed a valence electronic DP 
in the slab by performing a linear superposition of the valence 
pseudoatomic density calculated within  
the process of constructing the local ionic 
pseudopotential. We recall that the valence pseudoatomic density, which  
was obtained by means of a KS-type calculation, is a   
spherically symmetric function and in order to analize its role 
in setting the shape of the valence electronic DP, we have first  
integrated over its $x$ and $y$-dependencies; the resulting 
function $n_e(z)$, which is the one effectively used in 
the calculation of the valence electronic DP 
in the slab, is plotted for some representative 
systems, in figure \ref{nz-val-KMgSi}. For all systems, the corresponding   
$n_e(z)$ shows a decaying behavior whose width along with the 
associated weak Friedel oscillations may be identified as its 
basic characteristic features. 
The calculation of the valence electronic DP 
in the slab has proceeded as follows: by taking the 
OF-AIMD generated ionic positions we have located  
at each ionic position the previous $n_e(z)$ and we have 
carried out a configurational average. This approach is consistent with a 
linear response treatment of the valence electron density and therefore 
lacks any competition between kinetic and coulombic effects. 
The resulting valence 
electronic density profiles are compared with the OF-AIMD self-consistent 
ones for K, Mg and Si in figure \ref{denelecKMgSi}, which shows 
that both profiles are remarkably 
similar to each other. In particular, the phase of the valence 
electronic density oscillations is preserved, which suggests that the 
phase-shift must be determined by some feature embodied in the 
valence pseudoatomic density. 

In order to analize the different roles played by both features 
in modulating the phase-shift between the longitudinal 
 ionic and valence electronic DP, 
we have fitted the $n_e(z)$ to a model that shows no Friedel
oscillations but otherwise accurately reproduces its shape. We found 
that a rather good fit can be obtained for a model 
density of the normalized form $\exp[-|z/\sigma|^3]$, which 
includes just one parameter, $\sigma$.  
The specific values for the all the different systems are given in  
Table \ref{Results1} which shows a range of variation from 
$\sigma$ = $2.71$ \AA\ in Cs to $\sigma$ = $1.132$ \AA\ for Si.  
Furthermore, a comparison between the $n_e(z)$ and 
the above fitting model, is depicted in Figure \ref{nz-val-KMgSi} 
for K, Mg and Si. Besides the absence of the Friedel oscillations, it is 
observed that the model provides a rather good description 
of $n_e(z)$ for all the systems in this work. Now, by using the  
fitted model densities we have again performed its superposition according 
to the OF-AIMD generated ionic positions and the 
obtained valence electron profiles are shown 
in figure \ref{denelecKMgSi} for 
K, Mg and Si. Once again we observe that the 
corresponding valence electronic DP is 
virtually indistinguishable, for all systems, from that obtained by the 
superposition of valence pseudoatomic densities. More interestingly,  
the phase of the oscillations is preserved, and this fact suggests that 
the Friedel oscillations of the valence pseudodensity are irrelevant for this
question. Therefore, the reason for the different phase-shifts  
between ionic and valence electron oscillations must be found in the width of 
the valence pseudoatomic density (quantified by the parameter $\sigma$) as 
compared to the distance between layers in the profile (quantified by the 
wavelength of the longitudinal ionic oscillations $\lambda$).  
The ratio $\sigma$/$\lambda$ takes the values 0.62, 0.58 and 0.45 for 
K, Mg and Si respectively, which moreover correlates 
with a decreasing phase difference 
between their associated longitudinal ionic and valence electronic DPs. 
Table \ref{Results1} lists the obtained values of $\sigma$/$\lambda$ for all 
the systems considered in this work and three groups can be discerned: 
(i) the alkalis where $0.62 \le \sigma/\lambda \le 0.64$, (ii) Mg, Ba and Al 
with $0.55 \le \sigma/\lambda \le 0.59$, and (iii) Tl  and Si 
where $0.44 \le \sigma/\lambda \le 0.47$. Within each group, there is a 
similar phase difference between the corresponding longitudinal ionic and  
valence electronic DPs as epitomized by K, Mg and Si which may be 
considered as representatives for each group. 

\begin{figure}
\begin{center}
\mbox{\psfig{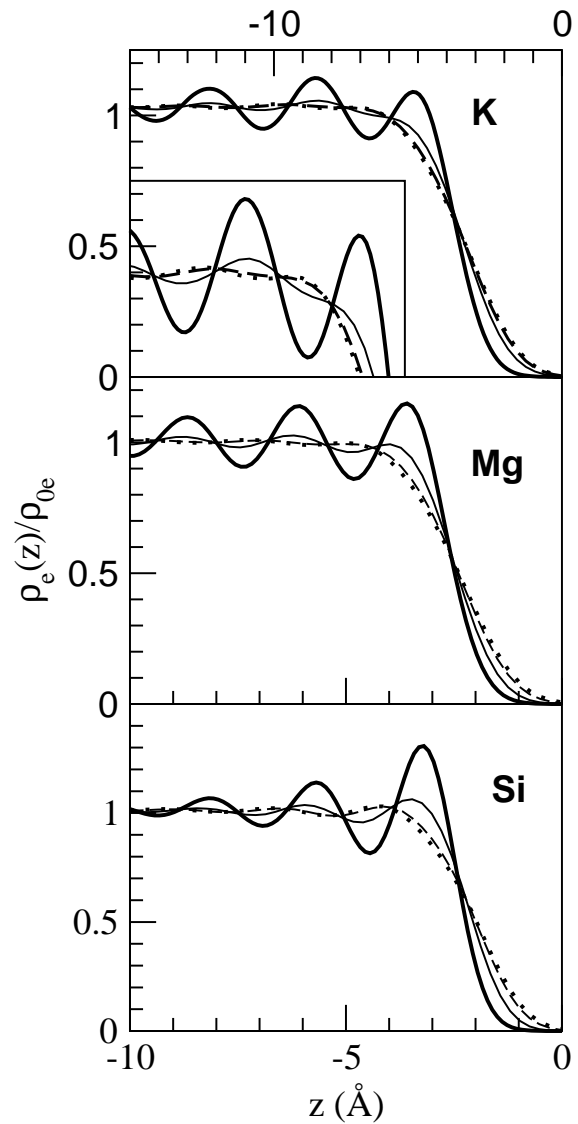}}
\end{center}
\caption{ Valence electronic density profiles obtained by superposition 
of model densities with different $\sigma/\lambda$ values, namely 
$\sigma/\lambda$= 0.45 (full line), 0.575 (broken line) and 
0.625 (dotted line). The thick line represents the longitudinal 
ionic density profile and 
the inset for K depicts the two outermost maxima.}
\label{models-kmgsi}
\end{figure}

As a further check on the previous argument, we have performed another 
test for the K, Mg and Si slabs. By taking their respective OF-AIMD 
generated ionic positions, we have again carried out a superposition 
of model densities with different widths, namely   
$\sigma/\lambda \approx$ 0.45, 0.575 and 0.625, which may be considered as 
representative values of the previous three groups. 
The calculated valence electronic DPs are depicted in 
figure \ref{models-kmgsi} which shows that, in the three systems,  
the resulting valence electronic DPs move from being nearly in-phase (when 
$\sigma/\lambda \approx$ 0.45) towards nearly opposite phase 
(when $\sigma/\lambda \approx$ 0.625) with respect to 
the corresponding longitudinal ionic DP. These results clearly showcase the 
crucial role played by the  
ratio $\sigma/\lambda$ in establishing the phase-shifts between the 
longitudinal ionic and valence electronic DPs. 
Moreover, the previous results provide a rationale for 
the aforementioned phase-shifts between the 
ionic and valence electronic DPs obtained for liquid Al in the present 
calculations and the OF-AIMD simulations of Jesson and 
Madden \cite{Jesson}. Specifically, our OF-AIMD results for 
the liquid-vapor interface of Al yielded $\lambda=2.35$ \AA, (with  
a ratio $\sigma/\lambda \approx 0.548$), whereas 
those of Jesson and Madden, which referred to liquid Al 
in contact with the (100) face of its solid fcc phase, 
gave $\lambda \approx 2.1$ \AA (which is close to the interlayer distance in 
the solid) and therefore leads to a $\sigma/\lambda \approx 0.613$. 
This latter ratio is close to the range obtained for the alkalis and 
therefore it foretells an opposite phase between 
the longitudinal ionic and valence electronic DPs, in accordance with 
Jesson and Madden's results. \cite{Jesson}    

\begin{figure}
\begin{center}
\mbox{\psfig{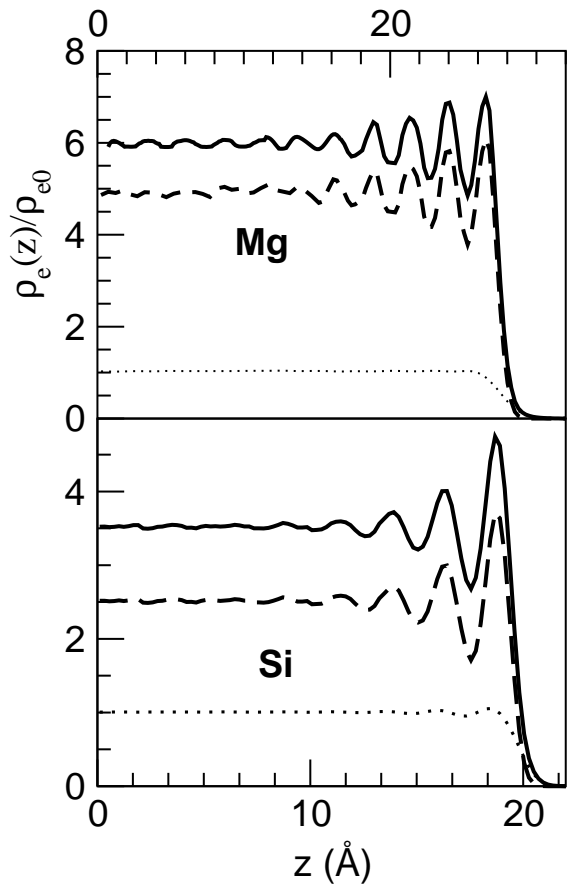}}
\end{center}
\caption{Valence (dotted line), core (dashed line) and total (full line) 
electronic density profiles (relative to the valence bulk value) for 
liquid Mg and Si. The longitudinal ionic DP is indistinguishable from the 
total electronic one.}
\label{totelec-mgsi}
\end{figure}

Experimental study of the LV interface is usually performed by  
X-ray reflectivity and/or grazing incidence X-ray diffraction techniques.  
Both methods measure the total electronic density profile which 
is the sum of both the core and valence electronic density contributions. 
Whereas the valence electronic DP is obtained by the OF-AIMD method, the 
corresponding core electronic DP has been calculated as follows. 
Using the OF-AIMD generated ionic positions we have placed 
at each ionic position the core electronic density (already computed in the 
process of calculating of the local pseudopotential) and we have 
carried out a configurational average. Once again, we note that the  
core electronic density is a spherically symmetric function 
which has previously been integrated over its $x$ and $y$-dependencies 
yielding a funcion $n_c(z)$, which is the one effectively used in 
the calculation of the core electronic DP. 
Figure \ref{totelec-mgsi} shows, for the Mg and Si slabs, a comparison 
among the obtained valence, core and total electronic DPs.  
Since the core electronic densities $n_c(z)$, 
are rather narrow funcions, their superposition leads to 
a core electronic DP which stands clearly in 
phase with the longitudinal ionic DP. Consequently, when it is 
added to the valence 
electronic DP it yields a total electronic DP whose 
phase-shift with respect to the longitudinal ionic DP 
depends on both the relative weight of the core electronic DP (always in 
phase) and the valence electronic DP (any phase is possible) as well as on 
the amplitude of the oscillations in both DPs. 
For the particular cases of Mg and Si, which are plotted in 
Figure \ref{totelec-mgsi}, their valence electronic DPs have relative 
weights of 1/6 and 2/7 respectively and, moreover,
the oscillations in their core electronic DPs are of 
substantially larger amplitude than those in their respective 
valence electronic DPs.  
Both features cooperate so as to produce a total electronic DP which 
practically coincides with longitudinal ionic DP. 
Indeed, this is a common trait in all the metallic systems considered in 
this work. Even liquid Li, which {\it a priori} could be more prone to 
changes because its valence electronic DP 
has a relative weight of 1/3 and also stands  
in opposite phase with the core electronic DP,  has a total electronic DP 
which closely matches the ionic one.

\section{Conclusions}

Results have been reported for the structure of the LV interface 
in several simple {\it sp-bonded} liquid metals. 
They have been obtained by an {\it ab initio} molecular dynamics 
method based on the 
density functional theory. Although it employs
an approximate electronic kinetic energy functional, 
the large variations in electron density associated with the 
liquid-vapour interface are accounted for selfconsistently
in the forces acting among the ions. The MD simulations 
were performed by using simulation slabs composed of 2000 ions, which 
are wide enough to discard possible interference effects between 
the two free surfaces. 

The calculated longitudinal ionic and electronic DPs exhibit clear 
oscillations, with the ionic ones lasting for at 
least four layers into the bulk liquid. The wavelength 
of the ionic oscillations shows good scaling with the radii of the associated 
Wigner-Seitz spheres; conversely no definite relationship with electronic 
parameters has been found. Furthermore,  
the metals with the greater valence display ionic DPs with 
the higher amplitude of the outer oscillation as well as a steeper 
decaying tail in the LV interface region.

Several structural properties were calculated, with special attention 
devoted towards its evolution along the outermost layers. Results were 
presented for the transverse pair distribution functions which in turn are 
used to analyze the atomic rearrangements occurring at LV interface. 
The coordination numbers remain practically constant for most of the slab and 
only very close to the LV interface there is a reduction
related to the absence of neighbors outside the interface. However, when 
compared properly, we have found that the angular distribution of the 
nearest neighbors doesn't change significantly across the interface. 
Moreover, it is shown that the interface induces weak structural 
rearrangements as compared with and ideally, step-like, terminated surface.

The valence electronic DPs show oscillations 
near the LV interface which are much weaker than those of the 
associated ionic DP. Moreover, it was found that the  
valence electronic DP is practically reproduced by superposing, at 
the ionic sites, the pseudoatomic valence densities calculated in 
the process of constructing the local pseudopotential. 
This suggests that for each system, the main features of its 
self-consistent valence electronic DP are already embodied in the 
characteristics of the corresponding pseudoatom valence density. 

We have also analized the mechanisms behind the relative phases of the 
oscillations in the ionic and valence electronic DPs.
It is found that those phases evolve    
from opposite phase in the alkalis to almost in-phase for Si. 
This is in stark contrast with the accepted wisdom, namely 
that the electronic and ionic DPs should oscillate in opposite phase.
An explanation is provided 
in terms of the size of the pseudoatoms (ion plus valence electronic 
cloud) relative to the distance between consecutive layers of the ionic 
profile.

\section*{Acknowledgments}

DJG and LEG acknowledge the financial support of the DGICYT of Spain 
(MAT2005-03415) and the EU FEDER program. 
MJS acknowledges the support of the NSERC of Canada.

\end{document}